\documentclass{aastex}
\usepackage{spr-astr-addons}
\usepackage{url}

\RequirePackage{color}

\begin{document}

\title{On Midrange Periodicities in Solar Radio Flux and Sunspot Areas}
\slugcomment{Not to appear in Nonlearned J., 45.}
\shorttitle{On Midrange Periodicities in Solar Radio Flux and Sunspot Areas}
\shortauthors{Mei et al.}

\author{Y. Mei\altaffilmark{1,2,3,4}} 
\and
\author{H. Deng\altaffilmark{1,2,3,4}}
\and
\author{F. Wang\altaffilmark{1,2,3,4}}
\email{fengwang@gzhu.edu.cn}


\altaffiltext{1}{Center For Astrophysics, Guangzhou University~510006, P.R.~China.}
\altaffiltext{2}{Yunnan Observatories, Chinese Academy of Sciences, Kunming~650216, P.R.~China.}
\altaffiltext{3}{CAS Key Laboratory of Solar Activity, National Astronomical Observatories, Beijing~100012, P.R.~China.}
\altaffiltext{4}{University of Chinese Academy of Sciences, Beijing~100049, P.R.~China.}

\begin{abstract}
Using the Hilbert-Huang transform technique, we investigate the midrange periodicities in solar radio flux at 2800 MHz (F10.7) and sunspot areas (SAs) from February 1, 1947 to September 30, 2016. The following prominent results are found: (1) The quasi-periodic oscillations of both data sets are not identical, such as the rotational cycle, the midrange periodicities, and the Schwabe cycle. In particular, the midrange periodicities ranging from 37.9 days to 297.3 days are related to the magnetic Rossby-type waves; 2) The 1.3-year and 1.7-year fluctuations in solar activity indicators are surface manifestations (from photosphere to corona) of magnetic flux changes generated deep inside the Sun; 3) At the timescale of the Schwabe cycle, \textbf{the complicated phase relationships} in the three intervals (1947-1958, 1959-1988, and 1989-2016) agree with the produced periodicities of the magnetic Rossby-type waves. \textbf{The findings indicate that the magnetic Rossby-type waves are the possible physical mechanism behind the midrange periodicities of solar activity indicators. Moreover, the significant change in the relationship between photospheric and coronal activity took place after the maximum of solar cycle 22 could be interpreted by the magnetic Rossby-type waves.}
\end{abstract}


\section{Introduction}

The temporal and spatial variations of solar magnetic activity have been studied extensively in the past few decades, but the nonlinear interaction among the Sun, the heliosphere, and the Earth has not been fully understood in detail, \textbf{partly} because of a large amount of complex variables \citep{2017LRSP...14....3U}. Long-term variability of solar magnetic activity could be studied through various indicators such as sunspot areas or numbers, polar faculae, coronal index, flare index, total solar irradiance, and so on \citep{2015LRSP...12....4H,2017JSWSC...7A..34D}. The availability of these databases has a potential function not only on solar physics researches but also on space weather and Earth's climate studies, as suggested by \cite{2014SSRv..186..105E}.


The database of sunspot areas (SAs) is the longest time series that was extensively applied to study the statistical properties of solar activity cycle. Solar radio flux (F10.7) observed at 10.7 cm or 2.8 GHz is a measurement of the integrated emission from whole sources of the Sun, it is directly related to the total amount of magnetic flux and results from magnetic resonance above sunspots and plages \citep{1990SoPh..127..321T,1994SoPh..150..305T}. \cite{1995MNRAS.274..858C} studied the physical relationship between solar UV flux and F10.7, and found that both of them could be \textbf{decomposed to form a basal component, a non-active component, and a part that is due to sunspot activity}. \textbf{SAs and F10.7 were found be highly correlated with each other as they are intrinsically inter-linked through solar magnetism and its dominant eleven year cycle.} However, \cite{2011ApJ...731...30K} concluded that F10.7, facular area, and maximum CME speed display better agreement with the large sunspot group numbers than they do with the small sunspot group numbers. \textbf{Solar activity indices actually represent one facet of solar activity, and the processes giving rise to that index are localized. That is, the various solar activity indices reflect the impact of the solar activity cycle
on different places and sets of processes in the Sun.}


The periodic variations of solar activity indicators have been observed with a wide range of timescales ranging from minutes to decades, and the most prominent oscillations are the 27-day rotational periodicity and the 11-year Schwabe cycle \citep{2009MNRAS.392.1159C,2013Ap&SS.343...27D}. The former reflects the modulation imposed on the solar flux at the Earth by solar rotation and the latter is related to the polarity reversal of solar magnetic fields. Based on various analysis techniques, many researchers studied the short-range and mid-range periodicities of  solar activity indicators following the discovery of a 154-day periodicity in solar flares \citep{1984Natur.312..623R}. The quasi-periodicities in the range of 1-4 years are referred to as quasi-biennial oscillations, and the periodicities shorter than 1 year are usually referred to as Reiger-type periodicities \citep{2003SoPh..212..201M,2010ApJ...709L...1V}. As pointed out by \cite{2014SSRv..186..359B}, the periodicities appear to be ubiquitous, being detected in magnetic activity indicators that are sensitive to the solar interior, and extending right out to the interplanetary medium. Therefore, the statistical analysis of solar magnetic activity indicators could provide information on the dynamical behaviors and the physical properties of the Sun, and could better understand the varying process of magnetic energy storage and dissipative mechanism. 

The values of sunspot areas could drop to zero during solar minimum, but the values of F10.7 do not, thus their relationship becomes more complex when the activity level of the Sun is very low, particularly during recent solar cycles. Whether or not the Sun is changing the dynamical behavior, and how important such transient changes might be the questions only addressable through examination of past data sets. Section 2 contains a brief description of the data sets and the time analysis approaches. Subsequently, the analysis results are revealed in Section 3. Finally we summarize the main findings and draw the conclusions in Section 4.

\section{Data Sets and Analysis Techniques}

\subsection{Observational Data}

The solar activity indices employed in this work are listed as follows:

(1).  The values of F10.7 are expresses in solar flux units (1 sfu = $\rm10^{-22}Wm^{-2}Hz^{-1}$) and have been systematically observed since February 1947 \citep{2013SpWea..11..394T,2014JApA...35....1B}. Both daily and monthly values of F10.7 are available from the Natural Resources Canada\footnote{http://www.spaceweather.gc.ca/solarflux/sx-en.php} and the National Geophysical Data Center\footnote{http://www.ngdc.noaa.gov/stp/spaceweather.html}. \textbf{Recent works \citep{2012SpWea..10.2011H,2015ApJ...808...29S} suggest that the F10.7 is a better proxy of the total magnetic flux on the solar surface than sunspot numbers.} This indicator displays a solar-cycle dependence \textbf{with emission exceed} during solar cycle maximum \citep{2009GeoRL..3610107D}, it thus constitutes the longest record of physical data of the magnetic activity variation. The data sets of F10.7 downloaded from the above websites are tabulated in two variants, the observed flux and the adjusted flux. The adjusted F10.7 has been calibrated for the varying Sun-Earth distance as if is were observed at 1 AU \citep{2011Ap&SS.332...73J}. Here, the adjusted values of daily F10.7 during the time interval from February 1, 1947 to September 30, 2016 are used for further analysis.

(2). SAs (in units of millionths of a solar hemisphere) are taken as proxies of the solar dynamo processes that are responsible for the establishment of large-scale magnetic fields \citep{2014ARA&A..52..251C}. By using measurements from photographic images obtained at the Royal Observatory in Greenwich (RGO) and some other observatories, the SAs were diligently recorded by the RGO from 1874 to 1976. From 1976 to present, the RGO measurements have been continued in the Debrecen Photoheliographic Data (DPD) sunspot catalogue that is compiled by the Debrecen Heliophysical Observatory, as commissioned by the International Astronomical Union (for details, please see the reviews wrote by \citealt{2014SSRv..186..105E} and \citealt{2015LRSP...12....4H}). The daily values of SAs from May 9, 1874 to September 30, 2016 are downloaded from the Marshall Space Flight Center (MSFC)'s website\footnote{http://solarscience.msfc.nasa.gov/greenwch.shtml}. The time series used in this study covers the time period February 1, 1947 to September 30, 2016, the common period to F10.7. In the models of total magnetic flux and total solar irradiance, SAs are often served as an important input parameter  \citep{2010JGRA..11512112K}, they thus possess more physical significance than sunspot numbers \citep{2013BASI...41..237F,2016AJ....151...70D,2010A&A...509A.100V}.

Figure 1 display the daily values of F10.7 (upper panel) and SAs (lower panel) for the time interval from February 1, 1947 to September 30, 2016. As the figure shown, both F10.7 and SAs wax and wane in 11-years Schwabe cycle. 

\begin{figure*}
	\centering
        \includegraphics[width=16cm]{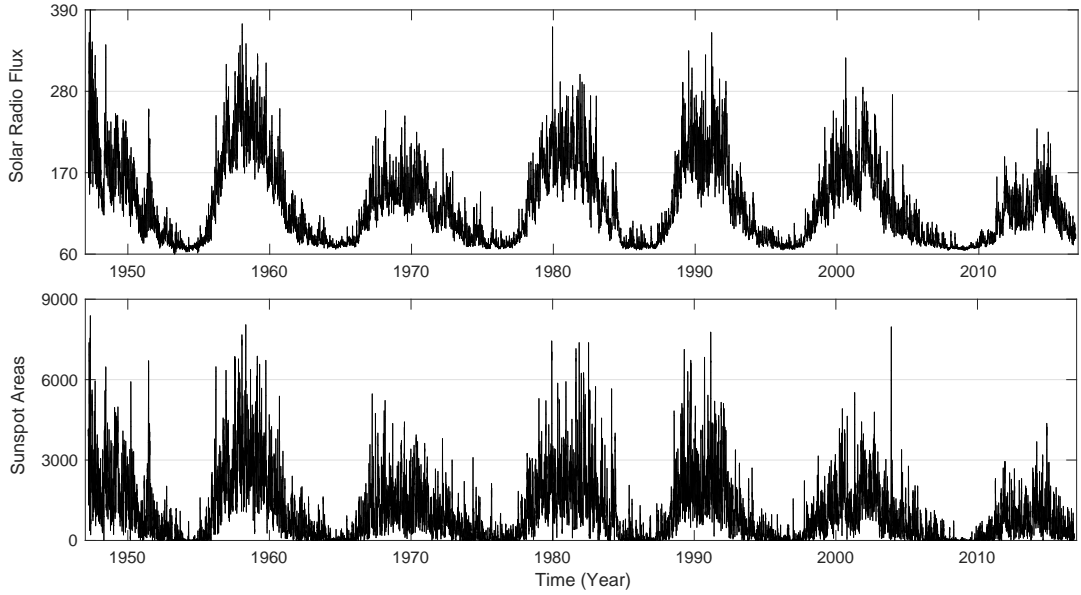}
\caption{Daily values of F10.7 (upper panel) and SAs (lower panel) for the time interval from February 1, 1947 to September 30, 2016.}
		\label{fig1}
\end{figure*}

\subsection{Hilbert-Huang Transform}
   
Hilbert-Huang transform (HHT), a fully data-driven approach introduced for nonlinear and non-stationary signal processing, is to assess the spectrum of the instantaneous frequency associated with the underlying systems \citep{1998RSPSA.454..903E,1999AnRFM..31..417H,2008RvGeo..46.2006H}. The key idea of HHT analysis is a combination of the empirical mode decomposition (EMD) to generate the intrinsic mode functions (IMFs) and the subsequent Hilbert spectral analysis (HAS) to identify the periodicities of extracted IMFs \citep{2014AdSpR..54..125D}. 

The goal of EMD technique is to decompose a given signal into a finite set of IMFs. The first extracted IMF represents the highest frequency component, and the lower frequency components correspond to higher order IMFs. Given a time series $s(t)$, it could be exactly reconstructed from IMFs using the following equation \citep{2004ApJ...614..435T,2016A&A...589A..56K}:    
\begin{equation}
     s(t)=
         \sum_{k=1}^{K}IMF_{k}(t)+r(t)
\end{equation}  
where $K$ is the total number of IMFs, $r(t)$ is the residue (either an adaptive trend or a constant) of the sifting process that represents the overall tendency of $s(t)$.

Once all of the IMFs are obtained, the HSA approach is applied to each IMF \citep{2013MNRAS.435.3639R}:
\begin{equation}
     H(IMF_{k}(t))=
         \frac {P}{\pi} \int_{-\infty}^{+\infty} \frac {IMF_{k}(\tau)}{t-\tau} d \tau
\end{equation} 
where $P$ denotes the Cauchy principle value of the singular integral. 

\textbf{To ensure that an IMF for EMD contains a true signal, we test the statistical significance of IMFs based on the method proposed by \cite{2004RSPSA.460.1597W}.}

\textbf{1). Calculate the energy of the IMFs. The energy of the $n$th IMF can be written as:}

\begin{equation}
     NE_n=
        \sum_{j=1}^{N}[C_n(j)]^2
\end{equation} 
\textbf{where $C_n(j)$ is the $n$th IMF and $N$ is the number of data points.}

\textbf{2). Ascertain any specific IMF contains little useful information, assume that the energy of that IMF comes solely from noise.}

\textbf{3). Use the energy level of that IMF to rescale the rest of IMFs.}

\textbf{4). If the energy level of any IMF lies above the theoretical reference white noise line, we can safely assume that this IMF contains statistically significant information at a selected confidence level ($e.g.,$ 95\% or 99\%). If the rescaled energy level lies below the theoretical white noise, then we can safely assume that the IMF contains little useful information.}


\begin{figure*}[t]
	\centering
        \includegraphics[width=16cm]{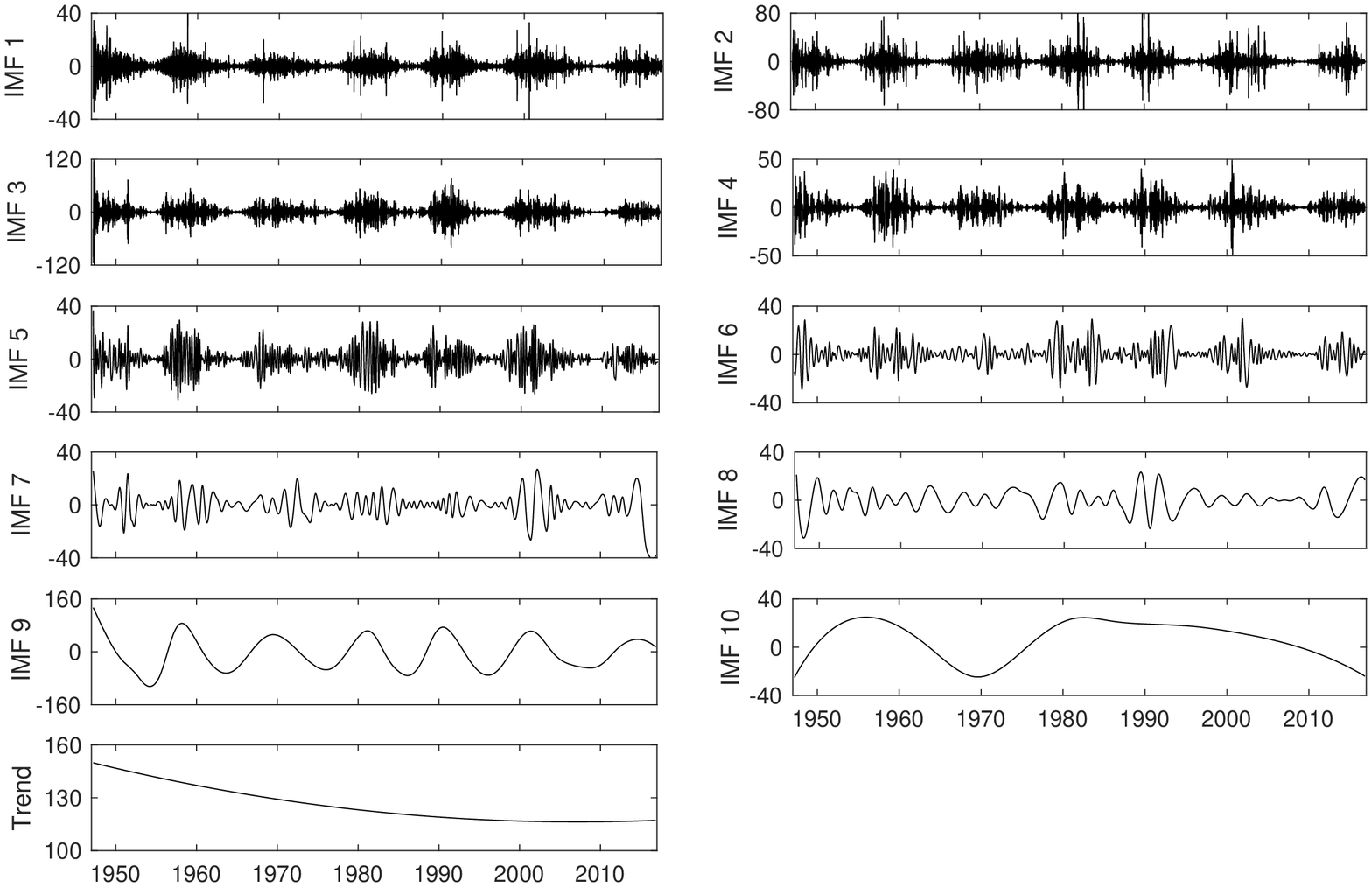}
        \caption{EEMD analysis of daily F10.7 from February 1, \textbf{1947} to September 30, 2016. It is decomposed into 10 IMFs and an adaptive trend (summed from IMF 11 to IMF 14).}
        \label{fig2}
\end{figure*}

\section{Results and Discussions}

Earlier papers showed that the major drawback of EMD analysis is the frequent appearance of mode mixing, leading to serious aliasing in the time-frequency distribution. To escape the mode-mixing problem, the noise-assisted ensemble EMD (EEMD) is applied to extract the daily time series of both F10.7 and SAs. EEMD technique consists of sifting an ensemble of white noise-added signal and treats the mean as the final true result, it is thus a substantial improvement over the EMD technique and solves the major drawback of mode mixing \citep{2016AJ....151...76X}.  \textbf{Here, both daily F10.7 and daily SAs are decomposed into 14 IMFs, but the last 4 IMFs give no significant periodicities. Therefore, the last 4 IMFs are amalgamated into a secular trend. Figures 2 and 3, respectively, display the 10 IMFs and the trend of daily F10.7 and daily SAs.} 

Subsequently, the HSA technique is applied to each of the extracted IMFs to reveal the average periodicities of both time series, and the results are shown in the left (daily F10.7) and right (daily SAs) panels of Figure 4, respectively. The solid and dashed lines denote the 99\% and 95\% confidence levels, respectively. \textbf{Here, the criteria that are used to determine the 99\% and 95\% confidence levels are based on the method proposed by \cite{2004RSPSA.460.1597W}. Actually, the problem of separating noise and signal is complicated and difficult when we do not know the level of noise in the solar data. As pointed out by \cite{2016A&A...592A.153K}, correctly accounting for the back-ground frequency-dependent random processes is certainly of a crucial importance when analyzing periodicities in solar activity indicators with EMD. We would like to point out that the applicability of the chi-squared law for both white and colored noises was justified by \cite{2004RSPSA.460.1597W} and \cite{2016A&A...592A.153K}, respectively. \cite{2017A&A...598L...2K} studied the quasi-periodic variations of the average magnetic field in a small-scale magnetic structure on the Sun. The EMD analysis of the original signal and the testing of the statistical significance of the intrinsic modes revealed the presence of the white and pink noisy components for the shorter-periods and the longer-periods of the spectrum, respectively. Therefore, the pink-noise criteria for EMD, obtained by \citep{2016A&A...592A.153K,2017A&A...598L...2K} are also suitable in this work.}

Table 1 collects the average periodicities of IMFs 1-10 for the two data sets. It is well known that the results of time-frequency analysis could be spurious, due to the truth that solar data sets are not fully stationary. As a well-known example, the Gleissberg cycle reported by many authors is found to be 80-100 years according to the techniques employed, but the length of the time series is very important. The detection of a periodicity of about 100 years, through classical spectral analysis (such as fast Fourier analysis) technique, requires a sample data of more than 400 years. Because the time interval of solar time series analyzed in this study is only 70 years, so the periodicities larger than 35 years are not taken into account.

\begin{figure*}[t]
	\centering
        \includegraphics[width=16cm]{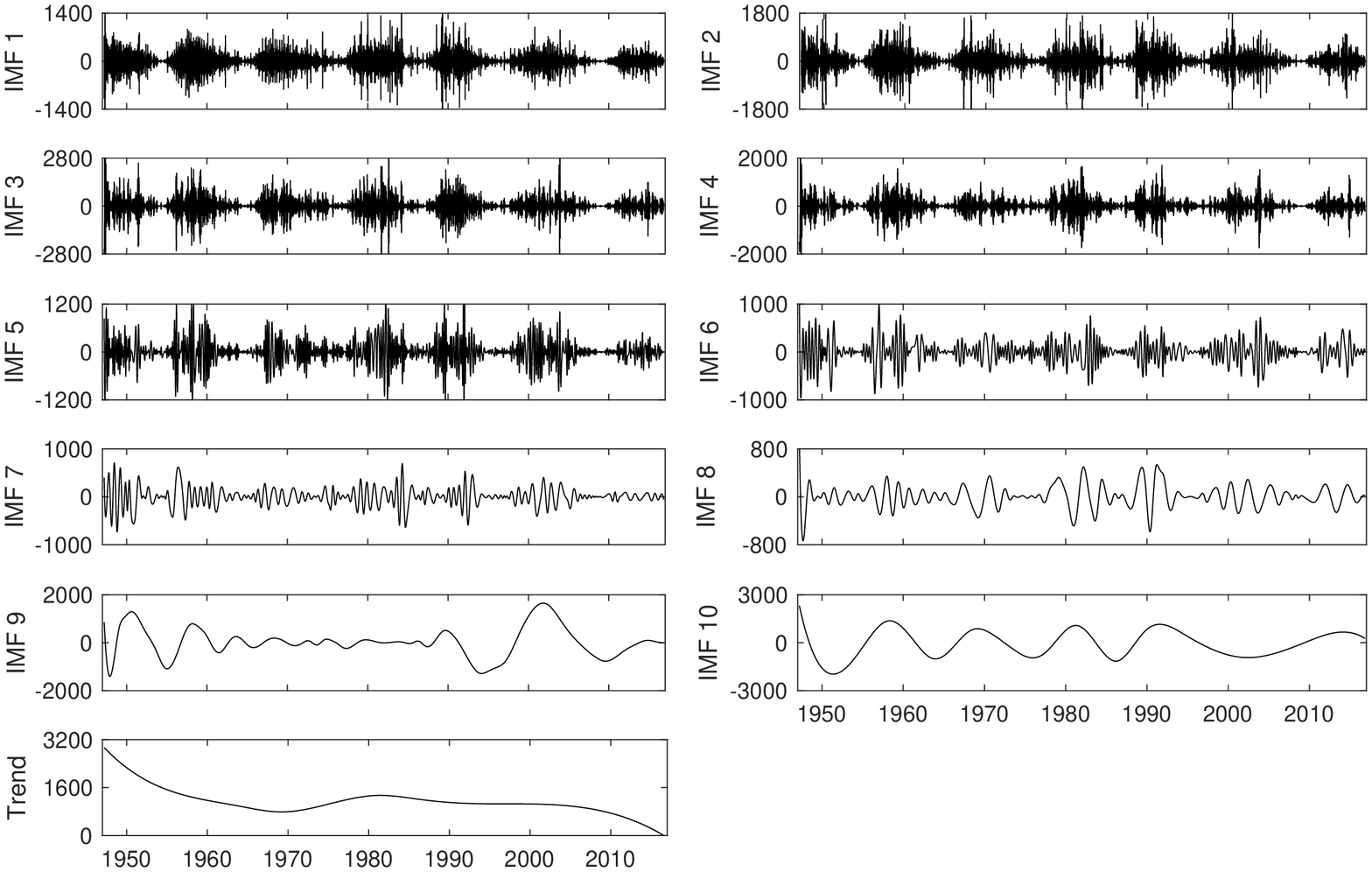}
        \caption{EEMD analysis of daily SAs from February 1, \textbf{1947} to September 30, 2016. It is decomposed into 10 IMFs and an adaptive trend (summed from IMF 11 to IMF 14).}
        \label{fig3}
\end{figure*}

\begin{figure*}[t]
	\centering
        \includegraphics[width=16cm]{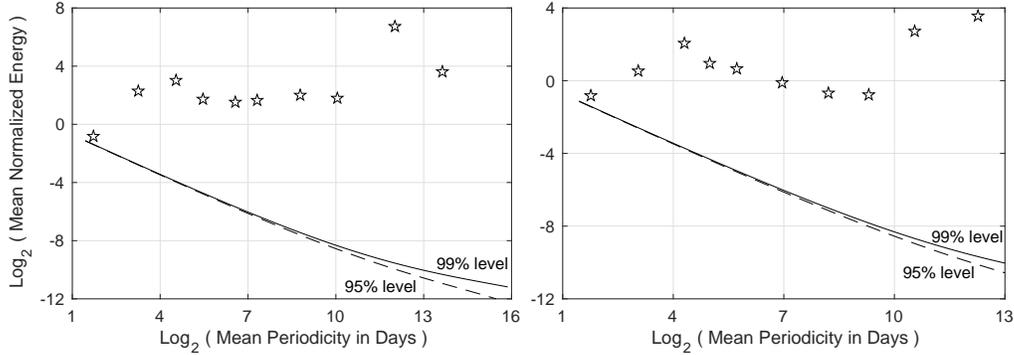}
        \caption{HSA of extracted IMFs from daily F10.7 (left panel) and daily SAs (right panel). Each sign represents the mean normalized energy of an IMF as a function of the mean periodicity of each IMF. The solid and dashed lines denote the 99\% and 95\% confidence levels, respectively.}
        \label{fig4}
\end{figure*}

   \begin{figure*}
   \centering
   \includegraphics[width=16cm]{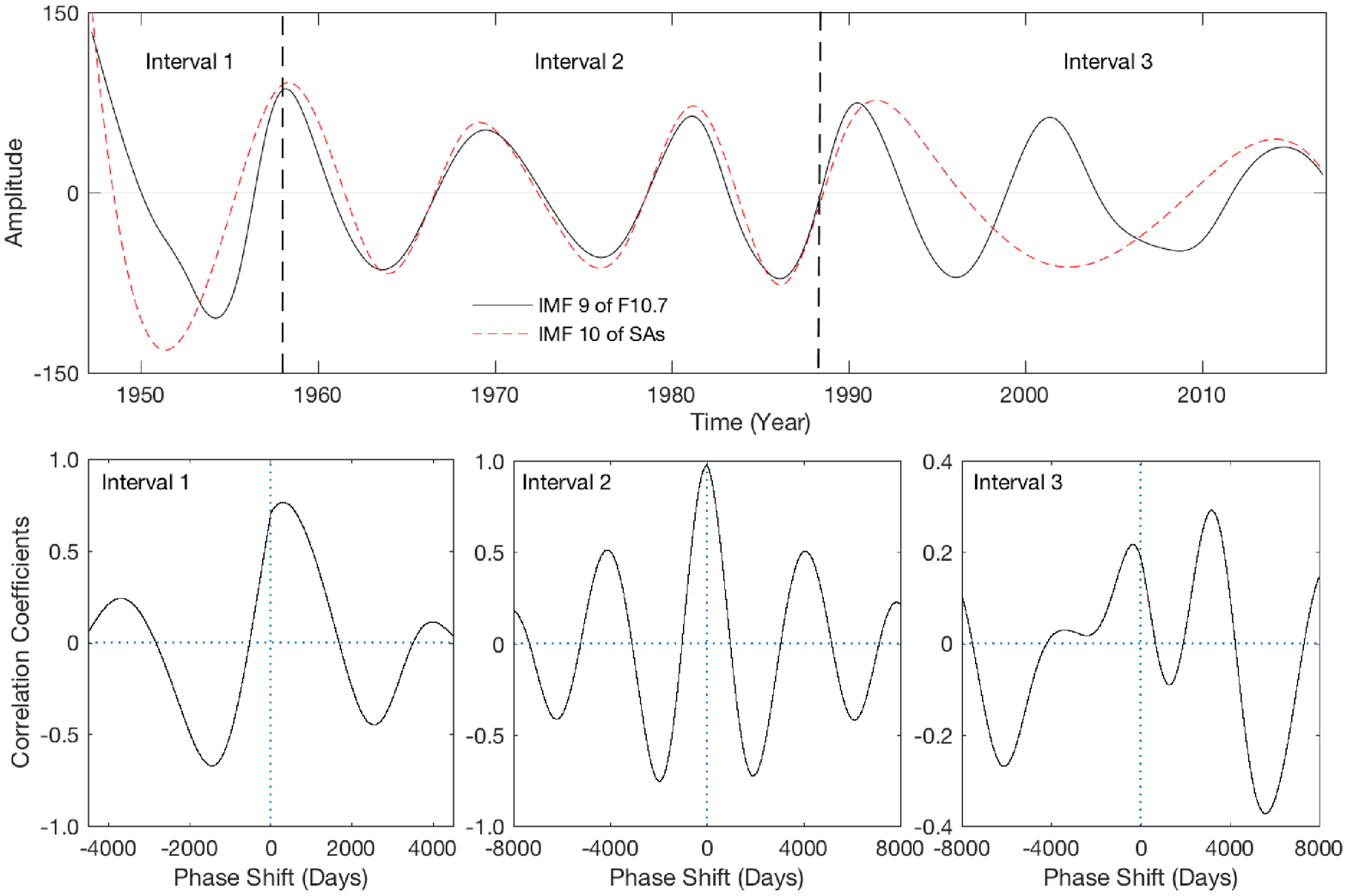}
   \caption{\small Upper panel: the Schwabe cycle components of daily F10.7 (IMF 9; black solid line) and daily SAs (IMF 10; red dotted line), respectively. Lower panels from left to right: cross-correlation coefficients between SCCF and SCCS for the interval I (1947-1958), interval II (1959-1988), and interval III (1989-2016), respectively.} 
   \label{fig5}
   \end{figure*}

In Table 1, the shortest periodicities (from 3 days to 10 days) in IMFs 1-2 for both data series are caused by the high-frequency noise and are not statistically significant. Perhaps the periodicity in IMF 10 for daily F10.7 are not veritable due to the limit length of the time series, but we cannot safe to say that it is not a real periodicity if the data length is longer in the future. The periods of 24.6 days in IMF 3 for daily F10.7 and 26.9 days in IMF 3 for daily SAs are the so-called rotational periodicity. \cite{2011MNRAS.414.3158C} studied the coronal rotation by analyzing daily F10.7 from 1947 to 2009, and found that the coronal rotation periodicity vary from 19 days to 29.5 days with an average of 24.3 days. \cite{2012RAA....12..187X} investigated the temporal variation of rotation cycle length of hemispheric sunspot numbers and found that their rotational cycle is about 27.4 days. \cite{2011Ap&SS.331..441L} applied the wavelet transform technique to investigate the temporal rotation cycle length of daily sunspot areas from May 9, 1874 to February 28, 2010, and found that the rotation periodicity is the only one at short time scales, and the value of the rotation periodicity is 27.4 days, at least from a global point of view.

\begin{table}
\caption{The average periodicities in IMFs 1-10 for daily F10.7 and daily SAs, respectively.
\label{table1}}
\begin{tabular}{p{18mm}p{25mm}p{25mm}}
\hline
                  & F10.7   & SAs   \\
\hline \hline
IMF1          & 3.3 d       & 3.3 d   \\
IMF2          & 9.5 d       & 8.3 d   \\
IMF3          & 24.6 d     & 26.9 d \\
IMF4          & 45.1 d     & 37.9 d \\
IMF5          & 96.8 d     & 65.4 d   \\
IMF6          & 165.1 d   & 125.5 d   \\
IMF7          & 1.21 yr    & 297.3 d \\
IMF8          & 2.91 yr    & 1.73 yr \\
IMF9          & 11.38 yr   & 3.44 yr   \\
IMF10        & 35.46 yr   & 13.53 yr \\
\hline
\end{tabular}
\end{table}

\begin{table}
\caption{The theoretical periodicities (in days) generated by magnetic Rossby-type waves according to the model proposed by \cite{2000ApJ...540.1102L}.
\label{table2}}
\begin{tabular}{p{12mm}p{12mm}p{12mm}p{12mm}p{12mm}}
\hline
n            & m       & P           & m       & P   \\
\hline \hline
01          & 01      & 25.35     & 16      & 201.6 \\
01          & 02      & 31.50     & 17      & 214.1 \\
01          & 03      & 41.92     & 18      & 226.6 \\
01          & 04      & 53.40     & 19      & 239.1 \\
01          & 05      & 65.31     & 20      & 251.6 \\
01          & 06      & 77.43     & 21      & 264.2 \\
01          & 07      & 89.68     & 22      & 276.7 \\
01          & 08      & 102.0     & 23      & 289.2 \\
01          & 09      & 114.4     & 24      & 301.7 \\
01          & 10      & 126.8     & 25      & 314.3 \\
01          & 11      & 139.2     & 26      & 326.8 \\
01          & 12      & 151.7     & 27      & 339.3 \\
01          & 13      & 164.1     & 28      & 351.9 \\
01          & 14      & 176.6     & 29      & 364.4 \\
01          & 15      & 189.1     & 30      & 376.9 \\
\hline
\end{tabular}
\end{table}

The periodicities in IMFs 4-7 of both time series, ranging from 37.9 to 297.3 days, are the so-called midrange periodicities. \textbf{Various solar activity indices, such as H$\alpha$ flare activity, photospheric magnetic flux, total solar irradiance, and coronal index, also have these midrange periodicities \citep{2006SoPh..233..319M,2010SoPh..264..255K,2011SoPh..270..365C}.} \cite{2003ApJ...591..406B} found these mid-term periodicities are very close to integral multiplies of the fundamental periodicity of 25.5 days. More importantly, the periodicity of 165.1 days in IMF 6 for daily F10.7 is very close to the Rieger-type periodicity (ranging from 150 days to 160 days), which was first revealed by \cite{1984Natur.312..623R} who studied the $\gamma$-ray flare data from Solar Maximum Mission. 

Several attempts were focused to the discovery of the physical source of Rieger-type periodicity, and many authors suggested that the possible mechanism behind this periodicity is the magnetic Rossby-type waves trapped in the solar surface \citep{2000ApJ...540.1102L,2008MNRAS.386.2278D,2010ApJ...709..749Z,2017ApJ...845...11F}. For typical solar parameters, the family of periodicities generated by the magnetic Rossby waves could be expressed as the following equation:
\begin{equation}
P_r\cong 25.1[m/2+0.17(2n+1)/m]. 
\end{equation}
where $m$ is an integer related to the wavenumber, and $n$ is an integer indicating the number of the considered nodes. If one assumes $n=1$, then the periodicities shown in Table 2 are yielded, depending on the assigned $m$ values. The reason for choosing $n=1$ is that it does not affect the resulting periodicities rigorously, as suggested by \cite{2000ApJ...540.1102L}. 

From Tables 1 and 2 one can see that the periodicities of 65.4 days, 125.5 days, 165.1 days, and 297.3 days are in better agreement with the produced periodicities of magnetic Rossby-type waves. At the same time, some other midrange periodicities of 37.9 days, 45.1 days, and 96.8 days are very close to the 41.9 days ($m=3$), and 102 days ($m=8$). That is to say, \textbf{we suggest} that the midrange periodicities of both F10.7 and SAs are related to magnetic Rossby-type waves. Actually, the unstable harmonics of magnetic Rossby-type waves lead to a periodic emergence of magnetic fluxes on the solar surface. Meanwhile, the growth rates of such waves link to the surface regions with strong magnetic fields. Most recently, \cite{2017NatAs...1E..86M} and \cite{2017NatAs...1E..96M} pointed out that magnetic Rossby-type waves are a type of global-scale wave that develops in planetary atmospheres, driven by the planet's rotation. These waves propagate westward owing to the Coriolis force, and their characterization enables more precise forecasting of weather on Earth. To better understand the physical origin of Rossby-type waves and their relationship with midrange periodicities of solar activity indices, more careful analyses of different data sets with different techniques are needed, as suggested by \cite{2016ApJ...826...55G}.

The periodicities of 1.21 years and 2.91 years in IMFs 7-8 for daily F10.7 are related to the 1.3-year periodicity, which exists in the solar wind oscillation and geomagnetic activity \citep{1994GeoRL..21.1559R, 2000AdSpR..25.1939M}, and solar internal rotation rate near the base of the convection layer \citep{2000Sci...287.2456H}. Meanwhile, the periods of 1.73 years and 3.44 years in IMFs 8-9 for SAs are inferred to be the 1.7-year periodicity, which is observed in cosmic ray intensity \citep{2003JGRA..108.1367K} and is considered as an important clue to understand the nature of solar magnetic cycle and magnetic flux emergence \citep{1996SoPh..167..409V}. These two periodicities have been determined from the interior of the Sun to the surface atmosphere, as well as from the interplanetary medium to the Earth's atmosphere. \cite{2014SoPh..289..707C} revisited these periodicities observed on the Sun, in the interplanetary space and Earth's magnetosphere. They expected to find the coupling behavior among the three spatially separated but linked regions in the heliosphere, and suggested a coherent relationship between the interplanetary space and the Earth's magnetosphere, but it is absent in the Sun. However, these two prominent periodicities are detected in green-line coronal index by \cite{2015JASTP.122...18D}. They suggested a scenario that the emergence and escape of magnetic flux from the convection zone, through the atmospheric layers and towards the heliospheric space, is a quasi-periodic process, with fluctuations in the coronal region playing a connective role in the Sun-heliosphere connection. Therefore, our findings lead to a conclusion that the 1.3-year and 1.7-year fluctuations in magnetic indicators are surface manifestations (from photosphere to corona) of changes in the magnetic fluxes generated deep inside the Sun.

The periodicities of 11.38 years in IMF 9 for daily F10.7 and 13.53 years in IMF 10 for daily SAs are the 11-years Schwabe cycle, but the average lengths for both time series are not identical. Here, these two IMFs are named as the Schwabe cycle component of F10.7 (SCCF) and SAs (SCCS) respectively, and we plot them in the upper panel of Figure 5 to show their differences. It is easy to see that the SCCS gets obviously larger with time, whereas the similar behavior is not obviously visible in the SCCF. Moreover, their phase relationship varies with time. To clearly display the phase relationship varying with time, the cross-correlation analysis is applied to calculated the correlation coefficients in three time intervals (interval I: 1947-1958; interval II: 1959-1988; interval III: 1989-2016), and the results are shown in the lower panels of Figure 5. \textbf{For example, for the time period 1959-1988 (interval II) the two seem in phase (the lower-middle panel of Figure 5), however they are not in phase during other intervals (intervals I and III; the lower-left and lower-right panels of Figure 5).} The abscissa in each panel indicates the leading and lagging shifts of SCCF with respect to SCCS along the calendar-time axis, with positive (negative) values representing forward (backward) shifts. For the interval I, when SCCF is shifted forward by 302 days, the cross-correlation coefficient reaches a peak value of \textbf{0.77}; for the interval II, the curve shows a roughly symmetric behavior between the leading and lagging shifts, and the largest value occurs at the -26 days with a correlation coefficient of \textbf{0.98}; the situation for the interval III is more complex than the former two,  when SCCF is shifted backward by -353 days, the correlation coefficient reaches a local maximum of \textbf{0.22}, and when SCCF is shifted forward by 3159 days, the correlation coefficient has a largest value of \textbf{0.29}, and the time interval between the two maxima is 3512 days. Here  \textbf{it should be pointed out} that all of the above correlation coefficients are statistically significant above the 95\% confidence level. Interestingly, the relative phase shifts in the three intervals, $i.e.,$ 302 days, -26 days, -353 days, and 3159 days (ten times of 315.9 days), are related to the family of periodicities (shown in Table 2) generated by the magnetic Rossby-type waves. Therefore, we infer that, at the time scale of Schwabe cycle, the magnetic Rossby-type waves could possibly be the reason for the complicated phase relationship between F10.7 and SAs. During the interval III, the activity level of the Sun is lower than that in the intervals I and II, so the phase relationship between SCCF and SCCA becomes more complex.

\section{Summary and Conclusion}

Long-term variations of solar magnetic indicators are very important for understanding and developing the solar dynamo models and their effects on space weather and Earth's climate. Using  the continuous time series of F10.7 and SAs, we studied the quasi-periodic oscillations of these two indicators. In the following, we highlight the important conclusions of this work: 

\begin{itemize}
\item The quasi-periodic oscillations of F10.7 and SAs are not identical, and the midrange periodicities ranging from 37.9 days to 297.3 days are related to the Rossby-type waves. 
\item The 1.3-year and 1.7-year fluctuations in solar activity indicators are surface manifestations (from photosphere to corona) of magnetic flux changes generated deep inside the Sun. 
\item For the component of the Schwabe cycle, their complicated phase relationships in different time intervals agree with the produced periodicities of the magnetic Rossby-type waves.
\end{itemize}

Solar magnetic indicators at different atmospheric layers are indications of the impact of solar activity cycle on more localized behaviors at different positions in the solar surface, the similarities and distinctions between F10.7 and SAs could improve our knowledge on the temporal variations of solar dynamical behaviors. The major origins of F10.7 have two components: one is the bremsstrahlung in coronal features which vary with magnetic network and plage/facular regions, and the other is gyro-magnetic radiation in active regions which vary with sunspot magnetic fields. F10.7 is preferred to describe the magnetic activity (both strong and weak fields) of the whole Sun to which the weak magnetic component mainly contribute. \textbf{The larger the sunspot areas are, the larger is the total magnetic field strength of sunspots. The strong relation between them makes the sunspot areas to the preferred index to represent the strong magnetic activity of the Sun (in sunspots), and thus the SAs could be used to represent the strong magnetic activity of the Sun.}  As both F10.7 and SAs are governed by the domain 11-year solar cycle, they exhibit many similar behaviors. However, their major magnetic contributions differ from each other, $i.e.,$ the weak magnetic fields to F10.7 while the strong magnetic fields to SAs, they thus display several different properties. In summary, solar magnetic activity indicators, from the lower atmosphere (photosphere and chromosphere) to the upper atmosphere (transition region and corona), are coupled in various styles of dynamical processes operating in the solar dynamo.

Magnetic Rossby-type waves are not only the probably physical mechanism behind the midrange periodicities of solar magnetic indicators, but are also the possible source for their complicated phase relationship at the component of Schwabe cycle. Since the year of 1989 (the maximum of solar cycle 22), the activity level of the Sun is lower than the previous years, which leads to the phase relationship between F10.7 and SAs becomes more complex. Meanwhile, a significant change in the relationship between activity indices in the photosphere and in the chromosphere/corona took place after the maximum of solar cycle 23 and continued into the current solar cycle 24 \citep{2017SoPh..292...73T}. \cite{2010ApJ...709..749Z} found that the magnetic Rossby-type waves are generally unstable and that the growth rates are sensitive to the magnetic field strength and to the latitudinal differential rotation parameters. Therefore, we infer that the changing behavior of the magnetic Rossby-type waves in the maximum of solar cycle 22 might be indicative of a relatively long-lasting minimum (from 2005-2010).
   
The changed behavior of solar magnetic activity during the nineties of 20th century might initiate the build-up to the one of the deepest solar minimum experienced in the past years. The analysis results presented in this work, and other unusual features (found by \citealt{2014SoPh..289...41B}, \citealt{2017MNRAS.470.1935H}, and so on) relating to cycles 23 and 24, could be the precursors to the long-lasting minimum and the dynamical processes of the solar dynamo. All of these results strengthen the earlier speculation that an obviously changing behavior of the Sun could be still in progress, and further studies are needed to focus on this aspect.

\acknowledgments
We would like to express our deep thanks to the staffs of the websites that provide the data sets and software codes for the public to download. Time series of solar radio flux are available from the Natural Resources Canada and the National Geophysical Data Center. Daily time series of sunspot areas are taken from the Royal Greenwich Observatory (RGO) archive and the Marshall Space Flight Center (MSFC)'s website. This work is supported by the National Key Research and Development Program of China (2016YFE0100300), the Joint Research Fund in Astronomy (No. U1531132, U1631129, U1231205) under cooperative agreement between the National Natural Science Foundation of China (NSFC) and the Chinese Academy of Sciences (CAS), the National Natural Science Foundation of China (No. 11403009 and 11463003), the open research program of CAS Key Laboratory of Solar Activity (KLSA201807), the Astronomical Big Data Joint Research Center, co-founded by National Astronomical Observatories, Chinese Academy of Sciences and Alibaba Cloud. \textbf{Last, but not least, the authors wish to express their gratitude to the anonymous referee and the editor for meaningful suggestions and comments that greatly improved the content and presentation of the paper.}

\bibliographystyle{aasjournal}
\bibliography{sts}

\end{document}